\documentclass[trackchanges,tighten,twocolumn]{aastex7}

\begin{document}

\title{Inhomogeneous stellar mixing in the final hours before the Cassiopeia A supernova}

\author[0000-0001-9267-1693]{Toshiki Sato}
\affiliation{Meiji University, Department of Physics}
\email[show]{toshiki@meiji.ac.jp}

\author[0009-0003-0653-2913]{Kai Matsunaga}
\affiliation{Kyoto University, Department of Physics}
\email{matsunaga.kai.i47@kyoto-u.jp}

\author[0000-0003-1518-2188]{Hiroyuki Uchida}
\affiliation{Kyoto University, Department of Physics}
\email{uchida@cr.scphys.kyoto-u.ac.jp}

\author[0000-0002-1104-7205]{Satoru Katsuda}
\affiliation{Saitama University, Department of Physics}
\email{katsuda@mail.saitama-u.ac.jp}

\author[0000-0002-6705-6303]{Koh Takahashi}
\affiliation{National Astronomical Observatory of Japan}
\email{koh.takahashi@nao.ac.jp}

\author{Hideyuki Umeda}
\affiliation{The University of Tokyo, Department of Astronomy}
\email{umeda@astron.s.u-tokyo.ac.jp}

\author[0000-0003-0304-9283]{Tomoya Takiwaki}
\affiliation{National Astronomical Observatory of Japan}
\email{takiwaki.tomoya.astro@gmail.com}

\author[0000-0003-4876-5996]{Ryo Sawada}
\affiliation{The University of Tokyo, Institute for Cosmic Ray Research}
\email{ryo@g.ecc.u-tokyo.ac.jp}

\author[0000-0002-5257-0617]{Takashi Yoshida}
\affiliation{Kyoto University, Yukawa Institute for Theoretical Physics}
\email{yoshida@yukawa.kyoto-u.ac.jp}

\author[0000-0002-8734-2147]{Ko Nakamura}
\affiliation{Fukuoka University, Department of Applied Physics}
\email{nakamurako@fukuoka-u.ac.jp}

\author{Yui Kuboike}
\affiliation{Meiji University, Department of Physics}
\email{ce257009@meiji.ac.jp}

\author[0000-0003-1415-5823]{Paul P. Plucinsky}
\affiliation{Center for Astrophysics — Harvard \& Smithsonian}
\email{pplucinsky@cfa.harvard.edu}

\author[0000-0002-8816-6800]{John P. Hughes}
\affiliation{Rutgers University, Department of Physics and Astronomy}
\email{jph@physics.rutgers.edu}

\begin{abstract}
Understanding stars and their evolution is a key goal of astronomical research and has long been a focus of human interest. In recent years, theorists have paid much attention to the final interior processes within massive stars, as they can be essential for revealing neutrino-driven supernova mechanisms and other potential transients of massive star collapse. However, it is challenging to observe directly the last hours of a massive star before explosion, since it is the supernova event that triggers the start of intense observational study. Here we report evidence for a final phase of stellar activity known as a ``shell merger'', an intense shell burning in which the O-burning shell swallows its outer C-/Ne-burning shell, deep within the progenitor's interior moments before the supernova explosion. In the violent convective layer created by the shell merger, Ne, which is abundant in the stellar O-rich layer, is burned as it is pulled inward, and Si, which is synthesized inside, is transported outward. 
The remnant still preserves some traces of such Ne-rich downflows and Si-rich upflows in the O-rich layer, suggesting that inhomogeneous shell-merger mixing began just hours ($\lesssim  10^4$ s) before its gravitational collapse. Our results provide the first observational evidence that the final stellar burning process rapidly alters the internal structure, leaving a pre-supernova asymmetry. This breaking of spherical symmetry facilitates the explosion of massive stars and influences various supernova and remnant characteristics, including explosion asymmetries and the neutron star's kick and spin.
\end{abstract}



\section{Introduction} 

The final stage of a massive star is known to be its most dramatic moment, with the massive Fe core of $\approx$1.5 solar masses ($M_\odot$) forming just hours to days before its gravitational collapse \citep[e.g.,][]{2002RvMP...74.1015W}. However there is still no way to observe this moment directly. After evolving into a supergiant, the core evolution of that star accelerates and cannot be distinguished on the Hertzsprung-Russell diagram. Even the nearby red supergiant Betelgeuse, which has been well studied for many years, makes it difficult to discuss how far its internal evolution has progressed \citep[e.g.,][]{2020ApJ...902...63J,2023MNRAS.526.2765S}. Our current understanding of the final stage of stellar evolution is based almost entirely on numerical simulations, and our ability to probe the actual interior conditions of dying stars is limited to difficult-to-observe signals such as neutrinos \citep[e.g.,][]{2015ApJ...808..168K,2016PhRvD..93l3012Y}.

\begin{figure}[t!]
 \begin{center}
  \includegraphics[bb=0 0 1110 1213,width=0.9\linewidth]{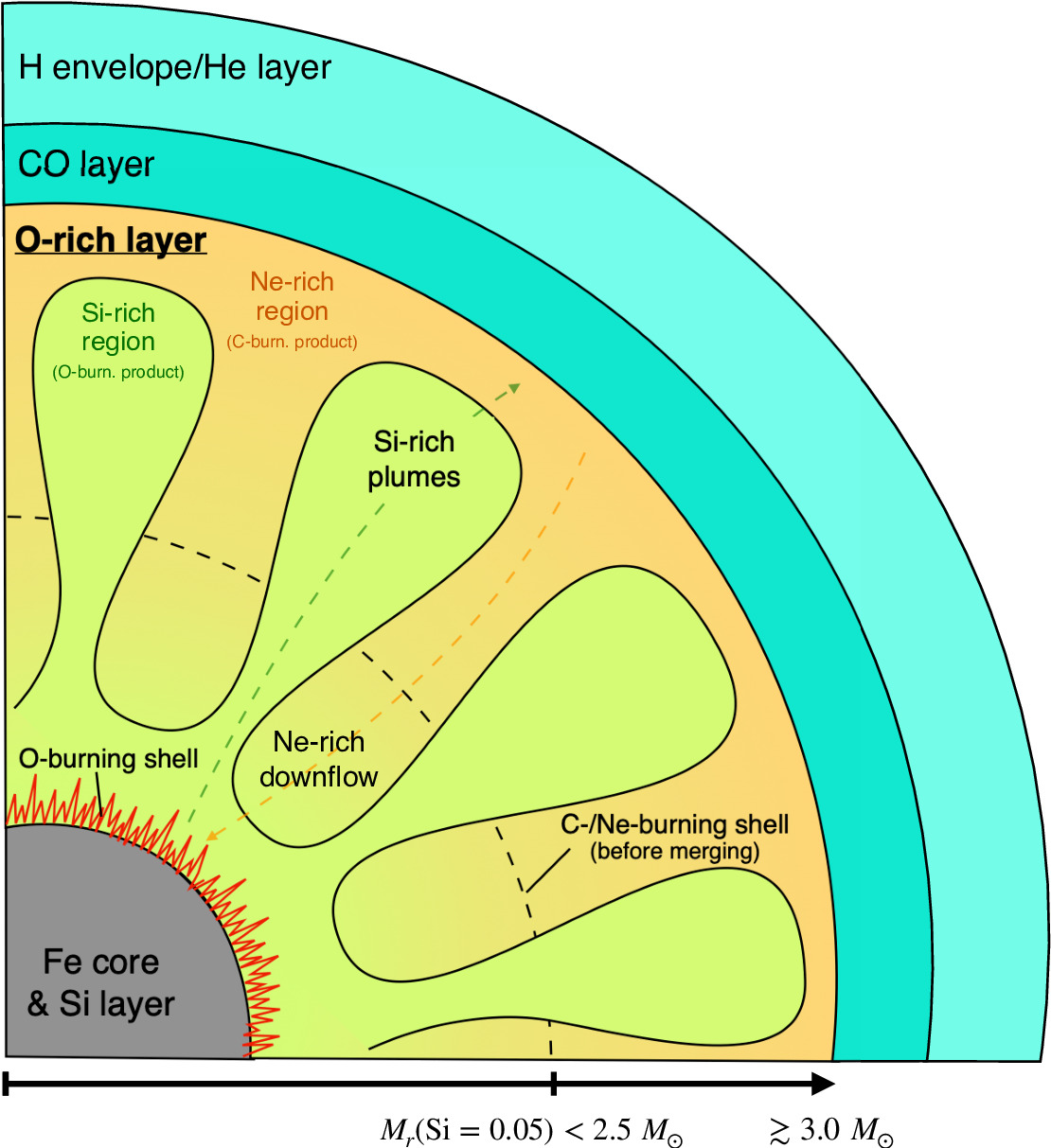}
 \end{center}
\caption{A sketch of a slice through the interior of a massive star in the process of a ``shell merger''. The shells are drawn roughly to scale by mass, but the plumes are a coarse representation of the expected complex convective ``fingers''. An intense O shell burning merges with the outer C-/-Ne-burning shell burning, forming a larger convective layer. In the merged layer, the extended convection efficiently brings Ne into the inner O-burning region (see the orange arrrow), and the amount of Ne is significantly decreased \citep{2020ApJ...890...94Y}. Inside such a progenitor, Si, which normally reaches up to about the bottom of the C-/Ne-burning shell, is transported further than that position (see the green arrrow). We define the Si mass radius $M_r$(Si=0.05), which is the maximum mass radius with the Si fraction greater than 0.05, to visualize the extension of the Si-rich region that is seen when shell mergers occur. In typical models that preserve the shell structures, the Si radius is less than $2.5~M_\odot$, but models that have experienced shell mergers indicate $\gtrsim 3.0 ~M_\odot$ (see also Figure \ref{fig:f2-new}). If the shell merger did not perfectly mix the materials within the O-rich layer, Ne-rich and Si-rich materials could be observed in the O-rich ejecta of its supernova.}
\label{fig:shell_pic}
\end{figure}

As for a massive star above about 10 $M_\odot$, Si core/shell burning and O-shell burning are thought to occur in a complex manner in the central region. Here, the O-burning shell burns strongly while supporting the outer part of the forming Fe core, and is crucial in determining the internal structure of the star before it collapses and explodes \citep[e.g.][]{1994ApJ...433L..41B,2014ApJ...783...10S}. An intense O-burning shell is predicted to occasionally break through to the outer C-/Ne-burning shell, forming one large convective layer (Figure \ref{fig:shell_pic}). Recent multi-dimensional simulations of supernova progenitors have shown a violent ``shell merger'' event \citep[][]{2011ApJ...733...78A,2016ApJ...833..124M,2019ApJ...881...16Y,2021MNRAS.506L..20Y,2024MNRAS.533..687R}, where the interface between the C-/Ne-burning shell (O-/Ne-rich layer) and the O-burning shell (Si-rich layer) disappears during evolution and Si is mixed far out into the O-/Ne-rich layer. During this process, Ne-rich material is inhomogeneously mixed downward into a Ne-poor region, where the entrained Ne is burned by incorporation into the inner O-burning region and its fraction decreases. As a result, a core-collapse supernova undergoing a shell merger is expected to have O-rich ejecta with low Ne/O and high Si/O ratios. This process generates the large-scale asphericities in the pre-supernova interior, promoting the growth of supernova asymmetry as seed perturbations \citep{2017MNRAS.472..491M,2018SSRv..214...33B,2020MNRAS.491..972A,2021ApJ...915...28B,2022MNRAS.510.4689V}. However, the large-scale mixing process with stellar asymmetry has not yet been confirmed by observations. 

In this study, we propose to probe pre–supernova interior asymmetries by examining the oxygen-rich ejecta of supernova remnants.  In particular, we target Cassiopeia A, one of the most famous asymmetric young remnants \citep{2000ApJ...528L.109H,2004NewAR..48...61V,2006ApJ...636..859F,2024ApJ...965L..27M,2024ApJ...976L...4D}, using deep Chandra ACIS-S spectroscopy to map key abundance ratios (Ne/Mg and Si/Mg) across multiple ejecta knots.  By comparing these observational diagnostics to predictions from theoretical models, we seek to determine whether the low-mode convective perturbations and compositional inhomogeneities generated in the final hours of stellar evolution survive collapse and shape remnant morphology.  

\section{Oxygen-rich ejecta in Cassiopeia A: their composition and distribution}
\begin{figure*}[t!]
\centering
\includegraphics[bb=0 0 1300 975, width=0.8\textwidth]{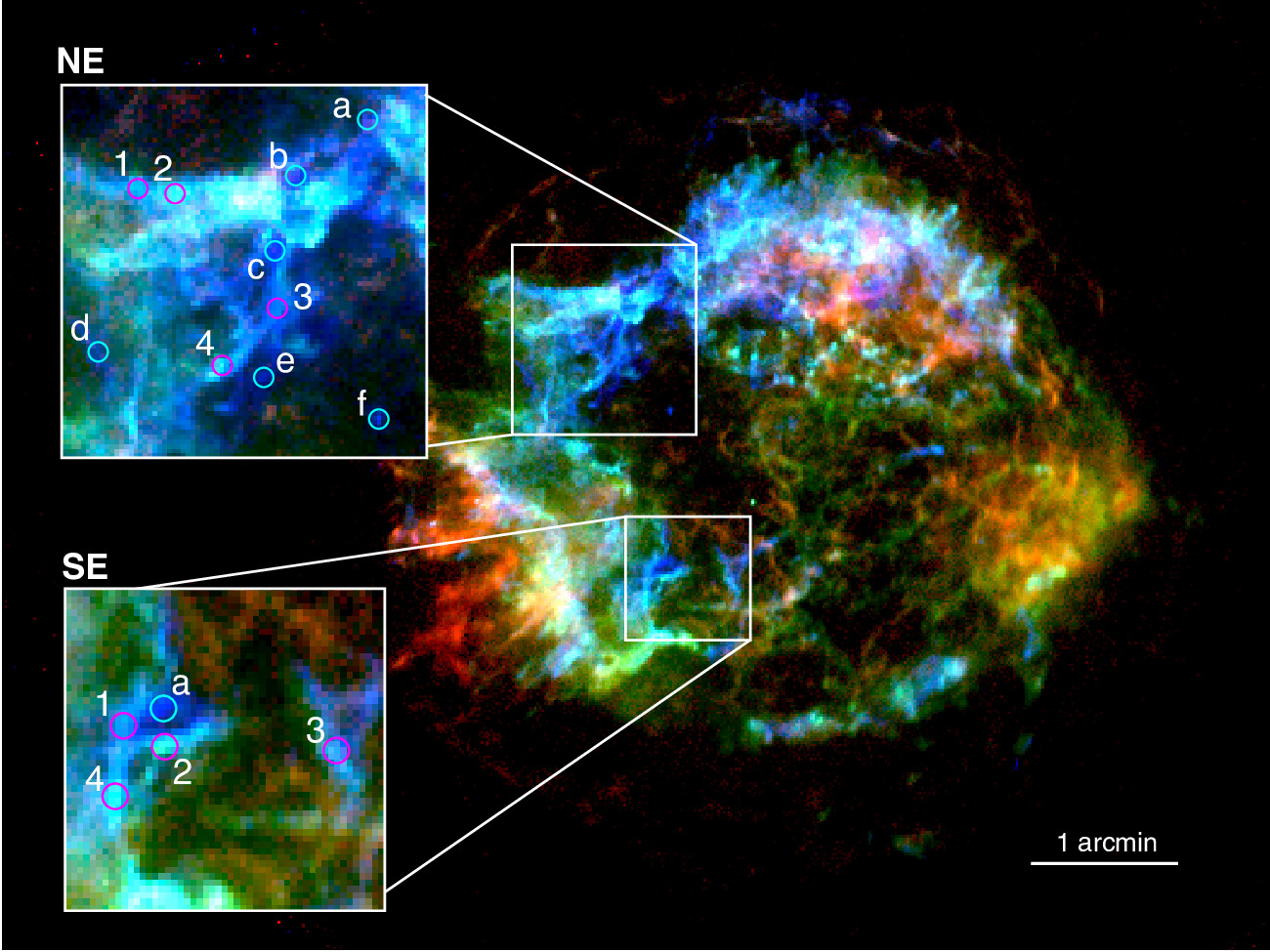}
\caption{Inhomogeneous elemental distribution in Cassiopeia A observed by Chandra. The difference in the mixing ratio of blue and green colors clearly shows the different composition in the O-rich ejecta, where red, green, and blue include emission within energy bands of 6.54--6.92 keV (Fe He$\alpha$), 1.76-1.94 keV (Si He$\alpha$), and 0.60-0.85 keV (O lines), respectively. The ejecta highlighted in red and green are products of explosive nucleosynthesis, while the ejecta in blue and emerald green reflect stellar nucleosynthesis. The circles in the small panels are O-rich regions used for spectral analysis. The regions of high and low X-ray intensity in the Si band are indicated by the magenta and cyan circles, respectively. Bright O-rich structures are excluded from the spectral analysis because of pile-up (see text).}\label{fig:f1}
\end{figure*}

\begin{figure*}[t!]
\centering
\includegraphics[bb=0 0 1638 730,width=1.0\textwidth]{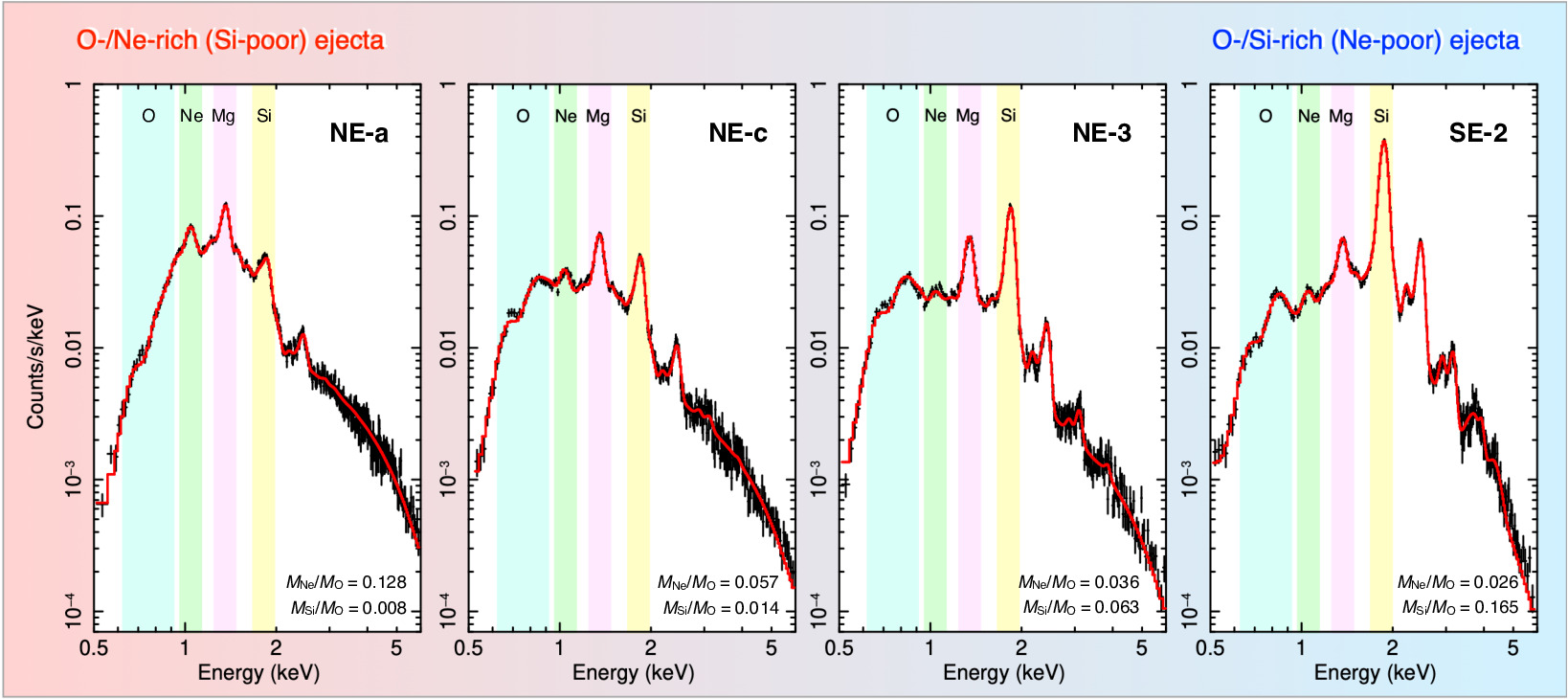}
\caption{A sequence of four X-ray spectra showing the gradual change of composition among O-rich ejecta regions in Cassiopeia A. From left to right, the compositions gradually change from O-/Ne-rich to O-/Si-rich. The solid red line shows the best-fit plasma model. The best-fit parameters of these fits are summarized in Table \ref{tab1}. Energy bands with characteristic elemental emission lines (O He$\alpha$+Ly$\alpha$+Ly$\beta$+Ly$\gamma$, Ne Ly$\alpha$, Mg He$\alpha$ and Si He$\alpha$) are highlighted in color. The regions (NE-a, NE-c, NE-3 and SE-2) were selected for easy viewing of the gradual change in elemental composition. The best-fit mass ratios (Ne/O and Si/O) are also shown in each panel.}\label{fig:S1}
\end{figure*}

The elemental composition of O-rich ejecta in a supernova, reflecting stellar hydrostatic nucleosynthesis, provides unique information about its progenitor interior. Cassiopeia~A is one of the best studied young SNRs with asymmetrically distributed O-rich ejecta \citep[e.g.,][]{2006ApJ...636..859F,2012ApJ...746..130H}. Since the 1970s, some observations have indicated that the remnant has low abundances of Ne and Mg \citep[e.g.,][]{1979ApJ...233..154C,1996A&A...307L..41V,2012ApJ...746..130H}, which are products in the stellar O-rich layer, but the origin of this peculiar composition is still not clear. %We here aim to search for O-rich ejecta in the remnant ejected from the extended convective layer formed by the O-C shell merger. 
The Ne/O ratio is about an order of magnitude lower than in the usual supernova models, and this trend is consistent with multi-wavelength observations with optical and X-rays \citep{1996A&A...307L..41V,1979ApJ...233..154C}. We found that progenitors that experienced shell mergers can reproduce the low Ne/O ratio well, while the others can not (see Appendix \ref{append1}). Thus, assuming a shell merger as the origin of the low Ne abundance in the remnant, we discuss the formation process of the peculiar O-rich ejecta of the remnant by comparing the elemental compositions with those in theoretical models. 

In this study, we focus on the asymmetric distribution of O-rich ejecta and their composition. Some simulations indicate the presence of a large-scale asymmetry in the shell-merger convective flow \citep[e.g.,][]{2020ApJ...890...94Y,2024MNRAS.533..687R}. Such inhomogeneities caused by shell mergers could be imprinted in the explosion geometry \citep{2017MNRAS.472..491M,2021ApJ...915...28B} and be observed in this asymmetric remnant \cite[e.g.,][]{2000ApJ...528L.109H,2015Sci...347..526M,2014Natur.506..339G,2021Natur.592..537S,2024ApJ...965L..27M}. 

\section{Observations and Data Analyis} \label{sec:obs}

Cassiopeia A has been observed several times since launch with Chandra's Advanced CCD Imaging Spectrometer (ACIS-S) X-ray imager \citep[e.g.,][]{2000ApJ...528L.109H,2004ApJ...615L.117H,2012ApJ...746..130H}. We used the deepest ACIS-S data targeting Cassiopeia A observed in 2004, with a total exposure of about 1 Ms. We reprocessed the event files (from level 1 to level 2) to remove pixel randomization and to correct for charge-coupled device (CCD) charge-transfer inefficiences using CIAO version 4.6 and CalDB 4.6.3. The bad grades were filtered out and good time intervals were accepted. 

Figure \ref{fig:f1} shows a three-color image of Cassiopeia A obtained by Chandra in 2004, where the Fe-rich, O-rich and Si-rich emissions are highlighted in red, blue and green, respectively. Since Fe is synthesized by explosive nucleosynthesis \citep[e.g.,][]{1995ApJS..101..181W}, we avoid Fe-rich (red) areas and focus on O- and Si-rich areas in the southeast (SE) and northeast (NE) regions. Interestingly, the image shows that the SE and NE regions are a mixture of O-rich, O-/Si-rich, and Si-rich areas, highlighted in blue, emerald green, and green, respectively. In particular, O and Si are expected to coexist in the O-rich layer of massive stars that have undergone a shell merger (Figure \ref{fig:S2}), but may remain inhomogeneous in some areas depending on mixing conditions up to gravitational collapse \citep[e.g.,][]{2020ApJ...890...94Y,2024MNRAS.533..687R}. Therefore, the different degrees of O and Si coexistence in the image may indicate the inhomogeneity due to its shell merger process. On the other hand, the mixing of elements after an explosion, as well as the propagation of the blast with Rayleigh-Taylor instabilities through a reverse shock, can also produce a similar situation \citep[e.g.,][]{2025ApJ...982....9V,2025A&A...696A.108O}. To understand whether this is the result of mixing in the progenitor star or mixing after the explosion, it is necessary to estimate the composition through detailed spectral analysis. To confirm the elemental compositions of O-rich areas where the degree of Si mixing appears to be non-uniform, we extracted X-ray spectra in the 0.5--6.0 keV band (Figure \ref{fig:S1}) from 15 regions defined in Figure \ref{fig:f1} and fitted them with an absorbed thermal plasma model (= {\tt phabs $\times$ vvpshock}) in Xspec using AtomDB version 3.0.9).

\begin{table*}[t]
\caption{Best-fit parameters of the spectral fits (single plasma model: {\tt phabs$\times$vvpshock}) in Fig.~\ref{fig:S1}. The solar
abundance (elemental number) ratios in \cite{1989GeCoA..53..197A} are used. The errors are 1$\sigma$ level.}\label{tab1}
\begin{tabular*}{\textwidth}{@{\extracolsep\fill}lcccc}
\toprule%
Region & NE-a & NE-c & NE-3 & SE-2 \\\hline
$N_{\rm H}$ [10$^{22}$ cm$^{-2}$]               & 1.39$^{+0.01}_{-0.02}$ & 1.15$\pm$0.01          & 1.10$^{+0.02}_{-0.01}$ & 1.18$\pm$0.01 \\
$kT_{\rm e}$ [keV]                              & 1.84$^{+0.03}_{-0.02}$ & 1.79$^{+0.03}_{-0.02}$ & 1.78$^{+0.03}_{-0.06}$ & 1.41$\pm$0.01 \\
(O/H)/(O/H)$_\odot$                                                & 4.9$^{+0.2}_{-0.5}$    & 6.3$^{+0.3}_{-0.6}$    & 4.7$^{+0.6}_{-0.4}$    & 6.1$^{+0.8}_{-0.1}$    \\
(Ne/H)/(Ne/H)$_\odot$                                               & 3.5$^{+0.1}_{-0.3}$    & 2.0$\pm$0.2            & 0.9$\pm$0.1    & 0.89$^{+0.02}_{-0.03}$    \\
(Mg/H)/(Mg/H)$_\odot$                                               & 1.81$^{+0.03}_{-0.08}$ & 1.63$^{+0.03}_{-0.04}$ & 1.63$^{+0.04}_{-0.21}$ & 1.70$\pm$0.03 \\
(Si/H)/(Si/H)$_\odot$                                               & 0.51$\pm$0.04          & 1.21$\pm$0.03          & 4.1$^{+0.4}_{-0.3}$ & 13.8$\pm$0.1          \\
(S/H)/(S/H)$_\odot$                                                & 0.41$\pm$0.03          & 0.94$^{+0.06}_{-0.05}$ & 2.4$^{+0.2}_{-0.1}$    & 10.9$^{+0.8}_{-0.1}$     \\
(Ar/H)/(Ar/H)$_\odot$  (=Ca)                                        & $<$0.13                & 0.5$\pm$0.3            & 3.1$\pm$0.4      & 10.2$\pm$0.5   \\
(Fe/H)/(Fe/H)$_\odot$  (=Ni)                                        & 0.16$^{+0.04}_{-0.05}$ & $<$0.06                & 0.29$\pm$0.04 & 0.15$\pm$0.01 \\
$n_{\rm e} t_{\rm low}$  [10$^{9}$ cm$^{-3}$ s] & 1.8$\pm$1.0            & 1.0$^{+1.4}_{-0.6}$    & 1.1$^{+0.7}_{-0.6}$ & 15.1$\pm$0.8            \\
$n_{\rm e}t_{\rm high}$  [10$^{10}$ cm$^{-3}$ s]& 7.7$^{+0.2}_{-0.1}$    & 6.1$^{+0.1}_{-0.2}$    & 3.93$^{+0.08}_{-0.07}$    & 8.4$\pm$0.1    \\
Redshift $z$ [$\times 10^{-2}$]                                  & $-$1.552$^{+0.005}_{-0.007}$    & $-$0.669$^{+0.005}_{-0.002}$    & $-$0.327$\pm$0.001    & $-$1.835$^{+0.007}_{-0.006}$   \\
Norm [$\frac{10^{-10}}{4\pi D^2}\int n_e n_H \mathrm{d}V$ cm$^{-5}$]  & 5.4$\pm$0.2    & 2.9$^{+0.2}_{-0.1}$    & 2.5$^{+0.2}_{-0.1}$    & 3.08$\pm$0.01    \\
$\chi^2$/d.o.f                                  & 364.21/299             & 364.73/262             & 419.60/241             & 436.89/267             \\
\hline
\end{tabular*}
\end{table*}

We here use small circular regions with radii of 2$^{\prime\prime}$ to avoid contamination from other areas as much as possible. Background spectra were extracted from the blank-sky region beyond the remnant. The O-rich emission from Cassiopeia A can be bright enough that significant pile-up of photon events can occur, distorting the spectra. Therefore we took care to select regions that avoided bright structures with pile-up. Fig.~\ref{fig:S1} shows four examples of X-ray spectra in the O-rich ejecta. From left to right, the elemental composition in the O-rich regions gradually changes from Ne-rich (Si-poor) to Si-rich (Ne-poor) with each mass ratio varying by nearly an order of magnitude. 

The increase in Si accompanying the decrease in Ne would be difficult to explain the effects of post-explosion mixing of O-rich and Si-rich materials. For example, if a supernova with an O-/Ne-rich layer that has not experienced a shell merger occurs, and Si-rich materials (O burning products) created by its explosive nucleosynthesis mix with the oxygen layer, this effect would not be observed. This is because the reduction in Ne cannot be explained simply by mixing O-/Ne-rich materials and Si-rich materials. On the other hand, as discussed in Appendix \ref{append1}, the abundance of Ne and Mg is extremely low throughout the remnant \citep{1996A&A...307L..41V,2012ApJ...746..130H}, supporting the shell merger hypothesis. If this shell merger perfectly mixed the O-rich layer, the O-/Ne-rich ejecta should not be observed. Thus, even if post-explosion mixing occurred in this case, the observed trend in elemental mass ratios cannot be reproduced. As a result, we conclude that the spectral variations in the O-rich ejecta originate from inhomogeneous mixing within the progenitor star. The best-fit parameters are summarized in Table~\ref{tab1} and the elemental mass ratios derived from the fitting results are summarized in Table~\ref{tab3}. We also used a different fitting function that included two thermal plasma models with abundances tied between them: {\tt phabs $\times$ (vvnei + vvnei)} (see Table~\ref{tab2} for the best-fit parameters). As we show below, our results are robust with respect to the choice of plasma model in our fits.

\begin{figure*}[t!]
\centering
\includegraphics[bb=0 0 1887 1564, width=1.0\textwidth]{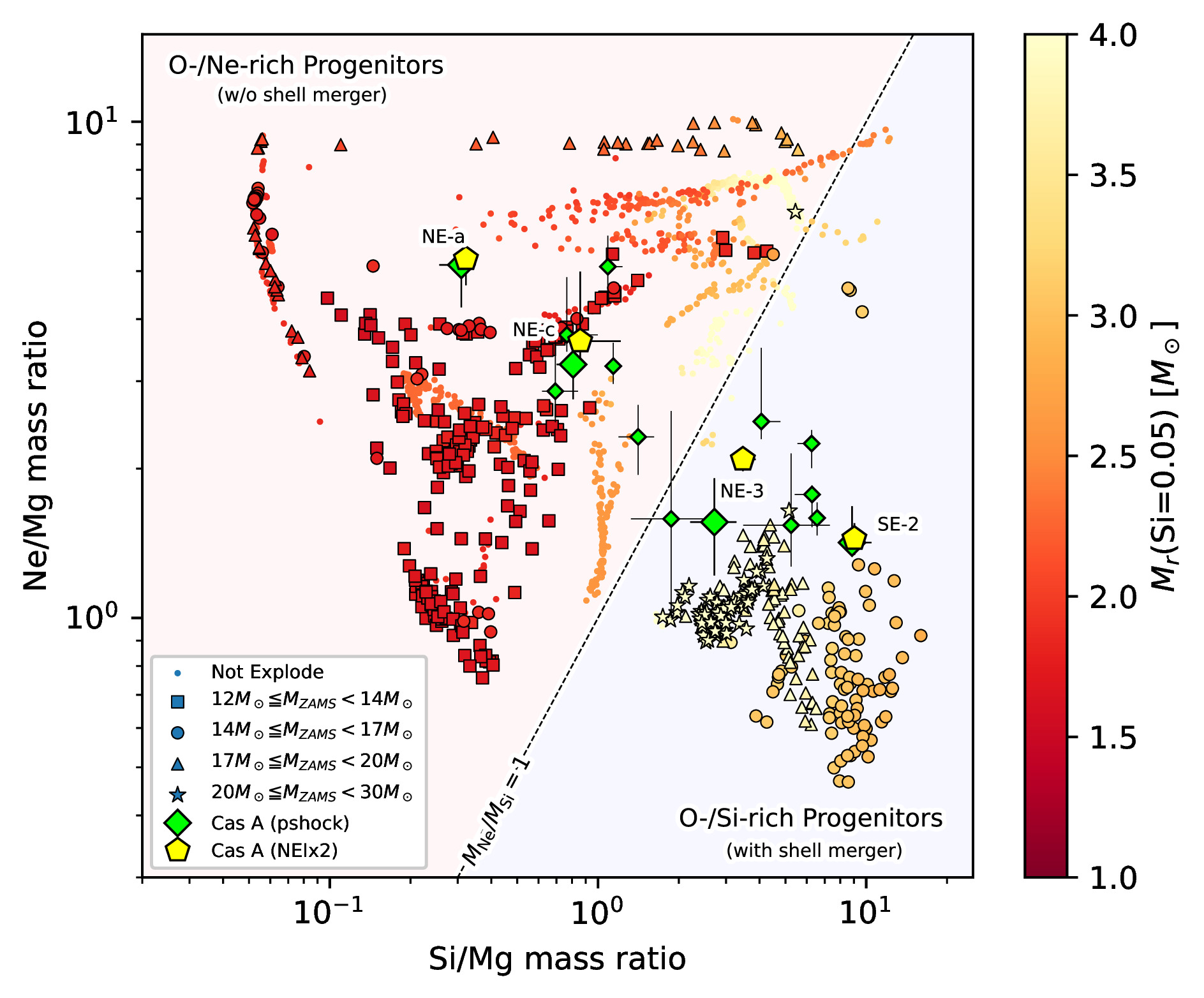}
\caption{Inhomogeneous shell-merger mixing suggested by Si/Mg and Ne/Mg ratios. The remnant retains characteristics of stars that underwent a shell merger (O-/Si-rich: lower right of panel) and those that did not (O-/Ne-rich: upper left of panel), suggesting that the shell merger did not completely mix the composition in the O-rich layer of the progenitor. The green diamonds show the observed mass ratios in fifteen O-rich regions using an absorbed single plasma model: {\tt phabs$\times$vvpshock} in Xspec. The large green diamonds show the results from four regions in Figure \ref{fig:S1}. The yellow pentagons show the same as the large green diamonds, but using different plasma model: {\tt phabs$\times$(vvnei+vvnei)}, which supports that the conclusion remains unchanged even when different plasma models are used. The colored data points are derived from 1D pre-supernova models in \cite{2018ApJ...860...93S}, where different symbols indicate different mass ranges. The radius of the Si-rich region in the progenitor, $M_r$(Si=0.05) (see Figure \ref{fig:shell_pic}), is shown in color. Using a BH–SN separation curve for the w18.0 calibration shown in \cite{2016ApJ...818..124E}, models that do not explode are identified and shown as colored dots.
}\label{fig:f2-new}
\end{figure*}

In our study, we defined the ``O-rich'' regions using Fig.~\ref{fig:f1}, where the X-ray flux of O emissions is large and that of Fe is small. We found that all the regions satisfy Fe/O $\lesssim 0.2$ (relative to the solar ratio). As seen in Table \ref{tab1}, most of the regions have Fe/O $<$ 0.1, thus we confirmed that these are indeed O-rich ejecta. The low Fe/O ratios also support a small contribution from the circumstellar (CSM)/interstellar medium (ISM).

\section{Inhomogeneous shell-merger mixing in the progenitor of Cassiopeia A}

In this section we relate our observational evidence for variation in the elemental composition of the O-rich ejecta in Cassiopeia A to the effects of an inhomogeneous merger of burning shells during the late evolutionary stages of the progenitor star.  We avoid using mass ratios to O in this comparison since O also exists outside the convection layer, which complicates theoretical predictions, and because the measurement of O emission is sensitive to systematic observational effects (see Appendix~\ref{append2} for more details).  Instead we focus our attention on the mass ratios of Ne and Si with respect to Mg, three species that coexist in the shell-merger convection region. The models we use are one-dimensional and cover a large range of individual star models that have or have not experienced shell mergers.

Figure~\ref{fig:f2-new} shows the clear negative correlation and strong variation from region to region of the Ne and Si mass fractions to Mg in the O-rich ejecta of the remnant. If the shell merger perfectly mixed the materials within the O-rich layer, then all of the ejecta should fall within the `O-/Si-rich Progenitors' region in the lower right of the figure. However, the O-rich ejecta of the remnant contains O-/Ne-rich materials, which supports an incomplete shell-merger mixing in which some of the Ne has survived (see Figure \ref{fig:shell_pic}). In Figure \ref{fig:f2-new}, the colored data points show theoretical predictions for the mass ratios. Here, we analyze 1,499 one-dimensional pre-supernova models (with the standard mass loss) provided by \cite{2018ApJ...860...93S} (see Appendix \ref{append3} for more details). This comprehensive dataset includes models that have undergone shell mergers, allowing us to extract the characteristics of this phenomenon during the final stages of pre-supernova evolution. The majority of the O-rich ejecta should have been synthesized during hydrostatic nucleosynthesis during stellar evolution, and its elemental composition is not so sensitive to explosive nucleosynthesis \citep[e.g.,][]{1995ApJS..101..181W,1996ApJ...460..408T}. Thus, we just calculated these mass ratios in the O-rich layer of pre-supernova models with a high oxygen mass fraction $>$ 0.4, assuming that the effects of explosive nucleosynthesis can be ignored \citep[see the discussions in][]{2024ApJ...970....4M,2025ApJ...984..185S}. To characterize shell-merger models, we also define the mass radius $M_r ({\rm Si}=0.05)$, which is the maximum mass radius with the Si fraction greater than 0.05 in a progenitor (see Figure \ref{fig:shell_pic} and Appendix \ref{append3}). In shell-merger models, strong O burning that penetrates and merges with the outer C-/Ne-burning shell creates a larger convective layer and carries Si further outward. Therefore, the model with an extended Si-rich region due to shell merger should have a larger $M_r ({\rm Si}=0.05)$. 

\begin{figure*}[t!]
\centering
\includegraphics[bb=0 0 2005 1224,width=1.0\textwidth]{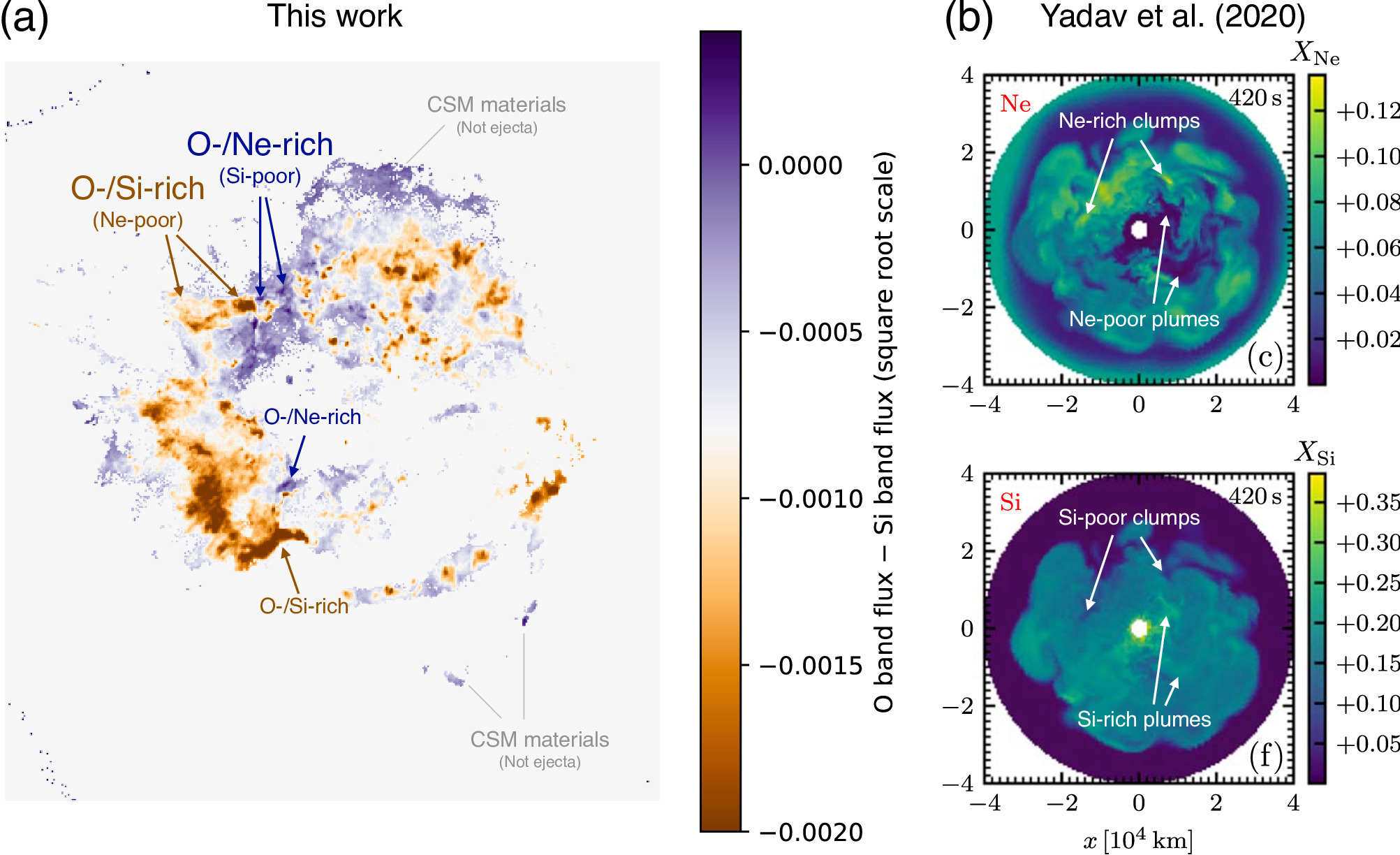}
\caption{A comparison of compositional variations in O-rich ejecta within Cassiopeia A and those in a 3D stellar simulation that underwent a shell merger \citep{2020ApJ...890...94Y}. (a) This image of Cassiopeia A was obtained by subtracting the Si band (1.76–1.94 keV) flux from the O band (0.6–0.85 keV) flux, measured by X-rays on a square-root scale. The figure was created using only O-rich areas (surface brightness $> 1.5\times10^{-7}$ photons s$^{-1}$ cm$^{-2}$). In this image, areas where O and Si coexist ($\approx$ O-/Si-rich ejecta) are highlighted in dark orange, while Si-poor areas ($\approx$ O-/Ne-rich ejecta) are highlighted in dark blue. (b) Image slice in the x-y plane of a 18.88 $M_\odot$ star model that underwent a shell merger \citep{2020ApJ...890...94Y}. The upper and lower panels show the mass fraction distributions of Ne and Si, respectively. Globally, the O-rich layer has a composition with low Ne and high Si, but clumpy small-scale structures that have avoided mixing remain.
}\label{fig:comp_Ydav}
\end{figure*}

In Figure \ref{fig:f2-new}, we found that models with $M_r ({\rm Si}=0.05) \gtrsim 3~M_\odot$ form an isolated group at the lower bottom of the Ne/Mg--Si/Mg plane, and most of them have experienced a shell merger. The orange circles seen in the shell-merger group are stars with $\sim$15~$M_\odot$, and because the oxygen layer is small in these stars, they can only carry Si up to around the mass radius of $\sim$3~$M_\odot$ (Figure \ref{fig:S2}). As for typical models with a standard O-/Ne-rich layer, the Si radius is significantly smaller, with $M_r ({\rm Si}=0.05) < 2.5~M_\odot$. As a result, two groups appear on this plot: `O-/Ne-rich Progenitors' that had not experienced shell merger and `O-/Si-rich Progenitors' that had. The O-rich ejecta of the remnant is distributed across these two groups, supporting the picture that the explosion occurred in a state that retained characteristics on both sides, or in other words, in a state of inhomogeneous shell-merger mixing. The fact that the shell merger was observed with the inhomogeneity remaining means that the explosion must occur before the convection has sufficiently mixed the stellar materials. \cite{2014ApJ...783...10S} found that shell mergers occur hours ($\sim10^4$ s) before the core collapse of 15--20 $M_\odot$ stars in their 1D calculations. However, in the case of 1D calculations, it is not possible to verify whether the inhomogeneity remains or not. In the case of 3D calculations, there are no examples of calculations being performed from the start of shell merger to the collapse, thus it is currently unknown to what extent inhomogeneity can be maintained. Because the convection velocity is faster in 3D simulations than in 1D simulations \citep[e.g.,][]{2020ApJ...890...94Y,2021MNRAS.506L..20Y,2024MNRAS.533..687R}, it may be necessary to consider a shell merger on a timescale shorter than $10^4$ s in order to explain the remaining inhomogeneity in Cassiopeia A. 

Shell mergers are known to produce global, low mode flow asymmetries dominated by $\ell\approx1$--$3$ modes \citep{2020ApJ...890...94Y,2024MNRAS.533..687R}. In contrast, the compositional inhomogeneities observed in the O-rich ejecta of Cassiopeia A in this work are of much smaller spatial extent compared to the global remnant morphology. Figure \ref{fig:comp_Ydav} shows a comparison of the spatial distributions of elements between our observations and the theoretical stellar model in \cite{2020ApJ...890...94Y}. We found that small-scale structures, like those seen in Cassiopeia A, can also be seen in 3D stellar models. As in Figure \ref{fig:comp_Ydav} (a), O-/Si-rich ejecta (i.e., dark orange structures in the figure) dominates throughout the entire remnant field, and shell-merger mixing appears to have homogenized most of the O-rich layers. On the other hand, O-rich and Si-poor ejecta (i.e., dark blue structures in the figure) exist in small clumpy shapes adjacent to the O-/Si-rich ejecta. Similarly, in Figure \ref{fig:comp_Ydav} (b), small-scale O-/Ne-rich structures surviving within global O-/Si-rich structures can be confirmed. Thus, while shell mergers may seed the global velocity asymmetries that drive explosion anisotropies, they could also imprint fine-scale compositional irregularities, resulting in the multi-scale asymmetry observed in supernova remnants. In the future, it will be necessary to compare the effects of mixing during the neutrino-driven explosion \citep[e.g.,][]{2025ApJ...982....9V} and the remnant phase \citep[e.g.,][]{2021A&A...645A..66O}. The mixing effects during the explosion and supernova remnant stage should greatly alter the spatial scale of asymmetric structures in the pre-supernova stage. However, the post-supernova mixing alone would not explain the inhomogeneity of the chemical composition in the remnant; it requires that the materials were unevenly mixed and burned in the progenitor.

\begin{figure*}[t!]
\centering
\includegraphics[bb=0 0 1878 1569,width=1.0\textwidth]{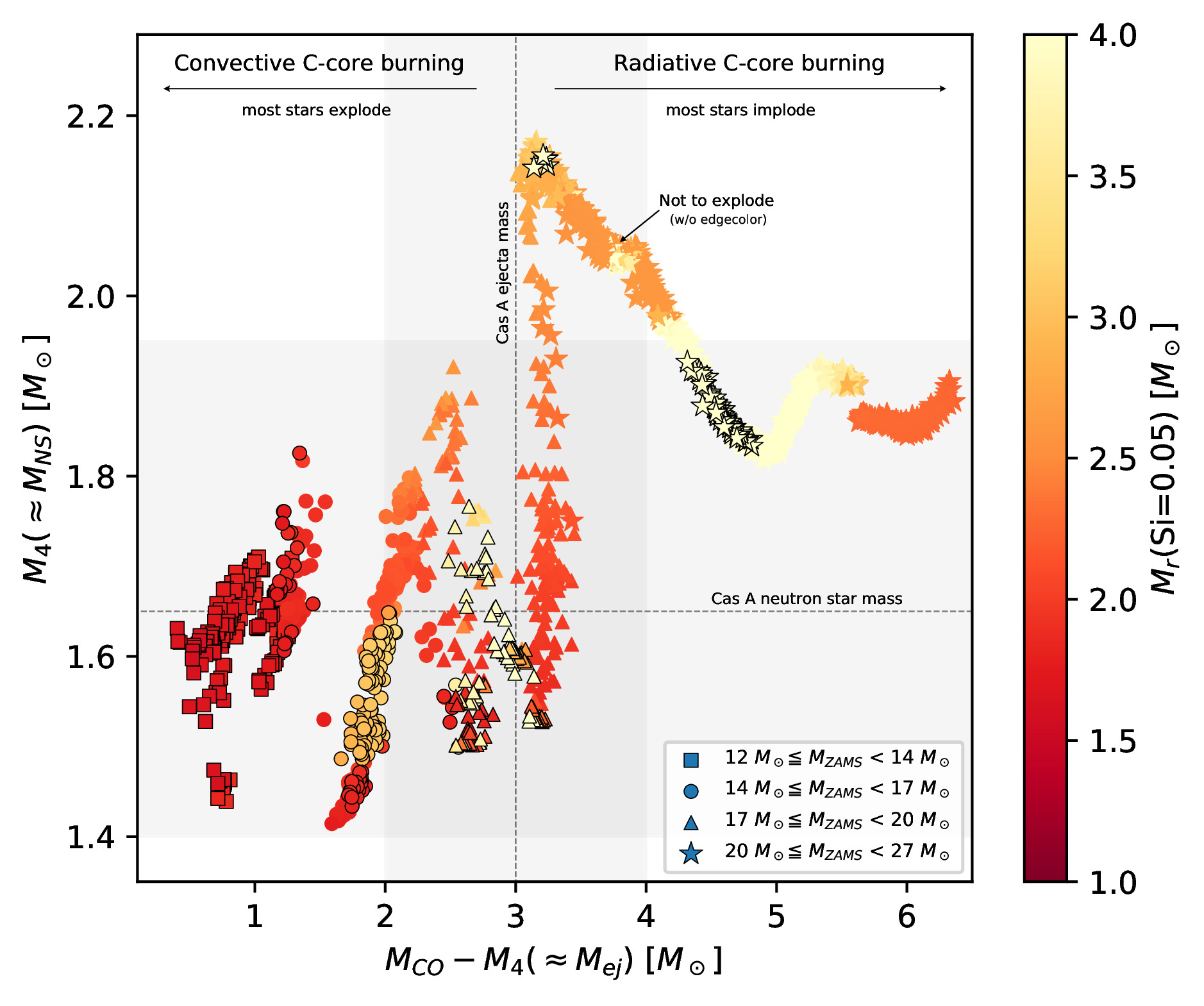}
\caption{Pre-supernova internal structure of Cassiopeia A inferred from its neutron star mass and ejecta mass. The vertical axis shows the core mass, $M_4$, enclosed at the point where the entropy in the progenitor exceeds a value of 4 $k_{\rm B}$ baryon$^{-1}$, which is similar to the neutron star mass after a supernova \citep{2016ApJ...818..124E,2018ApJ...860...93S}. The horizontal axis shows the mass of the CO core (the integrated mass up to the outermost oxygen layer) subtracted by $M_4$, which could be close to the total ejecta mass observed in X-rays. These parameters were calculated from the pre-supernova models in \cite{2018ApJ...860...93S}. The grey areas and broken lines in the vertical and horizontal directions are the estimated ejecta mass ($\approx3\pm1 ~ M_\odot$) and neutron star mass ($\approx1.65^{+0.30}_{-0.25}~M_\odot$) from observations \citep[][, respectively]{2012ApJ...746..130H,2011MNRAS.411.1977Y}. The color of the data points is the same as Figure \ref{fig:f2-new}. Using a BH–SN separation curve for the w18.0 calibration shown in \cite{2016ApJ...818..124E}, models that do not explode are identified and shown as data points without a black border.
}\label{fig:f4}
\end{figure*}

The shell-merger phenomenon has the potential to change the fate of a massive star in its final moments. For example, the strong turbulent flows created by the shell merger can promote the development of a supernova shock wave during the explosion \citep{2013ApJ...778L...7C,2015ApJ...799....5C,2017MNRAS.472..491M,2021Natur.589...29B,2021ApJ...915...28B}. In simulations, the probability of an explosion actually changes depending on whether realistic perturbations are present or not. Thus, the multi-dimensional stellar activity observed in this study could trigger the actual explosion of a star, and future multi-dimensional supernova simulations with more realistic pre-supernova structures will be able to clarify the effects in more detail. Furthermore, the shell merger is known to affect the formation of the stellar core, which is also dynamically important for the explosion \citep[e.g.,][]{2014ApJ...783...10S}. In the case of Cassiopeia~A, the observed ejecta mass of $\approx$3 $M_\odot$ and neutron star mass of $\approx$1.65 $M_\odot$ \citep{2012ApJ...746..130H,2011MNRAS.411.1977Y} suggest a progenitor of $\sim$17--20 $M_\odot$ for the remnant (Figure \ref{fig:f4}). In this mass range, there is a transition of the central C burning from the convective to the radiative regime \citep{2014ApJ...783...10S,2020ApJ...893...49S}. Radiative C-burning cores produce more massive O-burning cores, resulting in extended final core structures that are generally much more difficult to blow up. For example, we found that only $\sim$35\% of the 17--20 $M_\odot$ progenitors could explode at the Ertl criterion with the W18 engine \citep{2016ApJ...818..124E}, and $\sim$61\% of the explosive progenitors have experienced shell mergers. The mechanism that causes shell mergers is still being discussed  \citep{2018MNRAS.473.1695C,2025A&A...695A..71L} and needs to be studied in more detail in the future, but the observed properties of the remnant suggest an internal structure of a relatively massive progenitor ($\approx 18~M_\odot$), which is typically difficult to explode. On the other hand, the Ertl criterion does not always align with the results of numerical calculations \citep{2022MNRAS.517..543W,2024ApJ...964L..16B}. For example, supernova models with masses of 17--20 $M_\odot$ in \cite{2024ApJ...964L..16B} explode more energetically than others. Interestingly, previous studies have provided evidence for larger than nominal explosion energies in the range (1.5--4)$\times10^{51}$ erg for Cassiopeia A \citep{2003ApJ...593L..23C,2004NewAR..48...61V,2006ApJ...640..891Y,2016ApJ...822...22O,2017ApJ...842...13W,2020ApJ...893...49S}. 

The incomplete shell-merger mixing creates a pre-supernova asymmetry \citep{2020ApJ...890...94Y,2024MNRAS.533..687R} that grows during its gravitational collapse \citep{2000ApJ...535..402L,2014ApJ...794..162T} and leads to asymmetric shock propagation \citep{2017MNRAS.472..491M,2021ApJ...915...28B}. This can produce a highly asymmetric explosion with a strong neutron star kick \citep{2004ApJ...601L.175F,2017ApJ...837...84J,2024ApJ...964L..16B} as observed in Cassiopeia A \citep{2014Natur.506..339G,2021Natur.592..537S,2018ApJ...856...18K}. The asymmetric ejecta distribution can be preserved from immediately after the explosion to the present, several hundred years later \citep{2016ApJ...822...22O,2021A&A...645A..66O,2025A&A...696A.108O}. Thus, the inhomogeneity of the elemental composition within the progenitor could also be preserved to the present day. However, a full simulation of the entire process from the start of shell merger to gravitational collapse has not yet been carried out, and we are only just beginning to understand the evolution of supernovae into their remnants \citep[e.g.,][]{2020ApJ...888..111O,2021A&A...645A..66O}. To understand the details of the shell-merger mixing process and its effect on the supernova asymmetry, a future comparison between multi-dimensional simulations (from shell mergers to remnants) and observations is necessary.

Our model comparison is limited to non-rotating single stars, but the shell merger phenomenon itself can be strongly influenced by stellar rotation and binary interaction \citep{2021A&A...656A..58L,2024MNRAS.533..687R}. Therefore, the comparison between Cassiopeia A and the model should be revised in the future. In particular, it is possible that the outer layers of the progenitor of Cassiopeia A were stripped off by binary interaction \citep{2019A&A...623A..34K,2020ApJ...893...49S,2020MNRAS.499.1154H}, so the discussion based on the single star model may be insufficient. In addition, it appears that not only Ne but also Mg has decreased to a considerable extent. These observational facts will become important in considering the detailed physics behind them in the future.

\section{Summary and Conclusion}

We have presented the first direct observational evidence that a violent shell merger occurred in the progenitor of Cassiopeia A just hours before core collapse.  Deep Chandra ACIS‐S spectroscopy of fifteen spatially distinct O‐rich ejecta regions reveals mass‐ratio variations (Ne/Mg and Si/Mg) spanning nearly an order of magnitude.  These variations trace the incomplete mixing predicted by recent multi‐dimensional shell‐merger models.  

The coexistence of compact ejecta regions in both the ‘O-/Ne‐rich’ and ‘O-/Si‐rich’ regimes implies that the merger did not fully homogenize the O‐rich layer prior to collapse, leaving behind multi‐scale compositional inhomogeneities and asymmetric velocity fields.  Such pre‐supernova perturbations are expected to seed explosion asymmetries and aid shock revival. Future work should combine high‐precision X‐ray spectroscopy (e.g., Athena, XRISM) with fully three‐dimensional progenitor and explosion simulations that resolve shell-merger convection and track its inhomogeneity through collapse and into the remnant phase. Such efforts will be crucial to quantify the impact of shell mergers on supernova asymmetries, neutron‐star kicks, and remnant evolution.

For a long time in the history of astronomy, it has been a dream to study the internal structure of stars. Through this research, we have gained insight into the last hours ($\lesssim10^4$ s) of the interior of a massive star with lifetimes of $\sim10^6$--$10^7$ years. This moment not only has a significant impact on the fate of a star, but also creates a more asymmetric supernova explosion.

\begin{acknowledgments}
We are grateful to David Vartanyan for valuable discussion. This work was supported by the Japan Society for the Promotion of Science (JSPS) KAKENHI grant No. JP23K13128. J.P.H., the George A. and Margaret M. Downsbrough Professor of Astrophysics, acknowledges the Downsbrough heirs and the estate of George Atha Downsbrough for their support. His Chandra research on supernova remnants was additionally supported by SAO award number G01-22058X. We also thank the anonymous referee for comments that helped us improve the manuscript.
\end{acknowledgments}

\appendix

\renewcommand{\thefigure}{A\arabic{figure}}
\renewcommand{\thetable}{A\arabic{table}}

\begin{figure*}[h!]
\centering
\includegraphics[bb=0 0 1916 1564,width=0.495\textwidth]{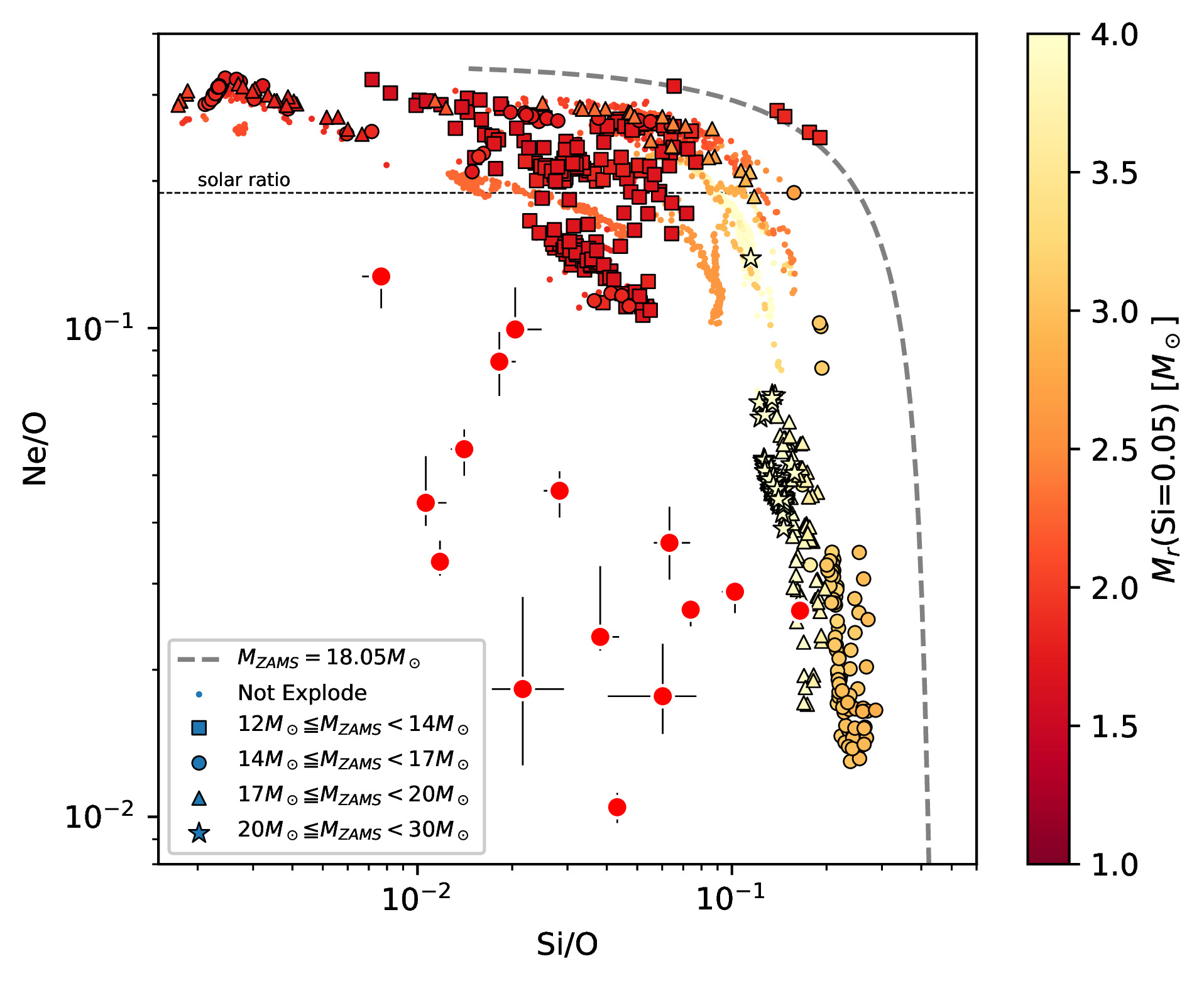}
\includegraphics[bb=0 0 1916 1564,width=0.495\textwidth]{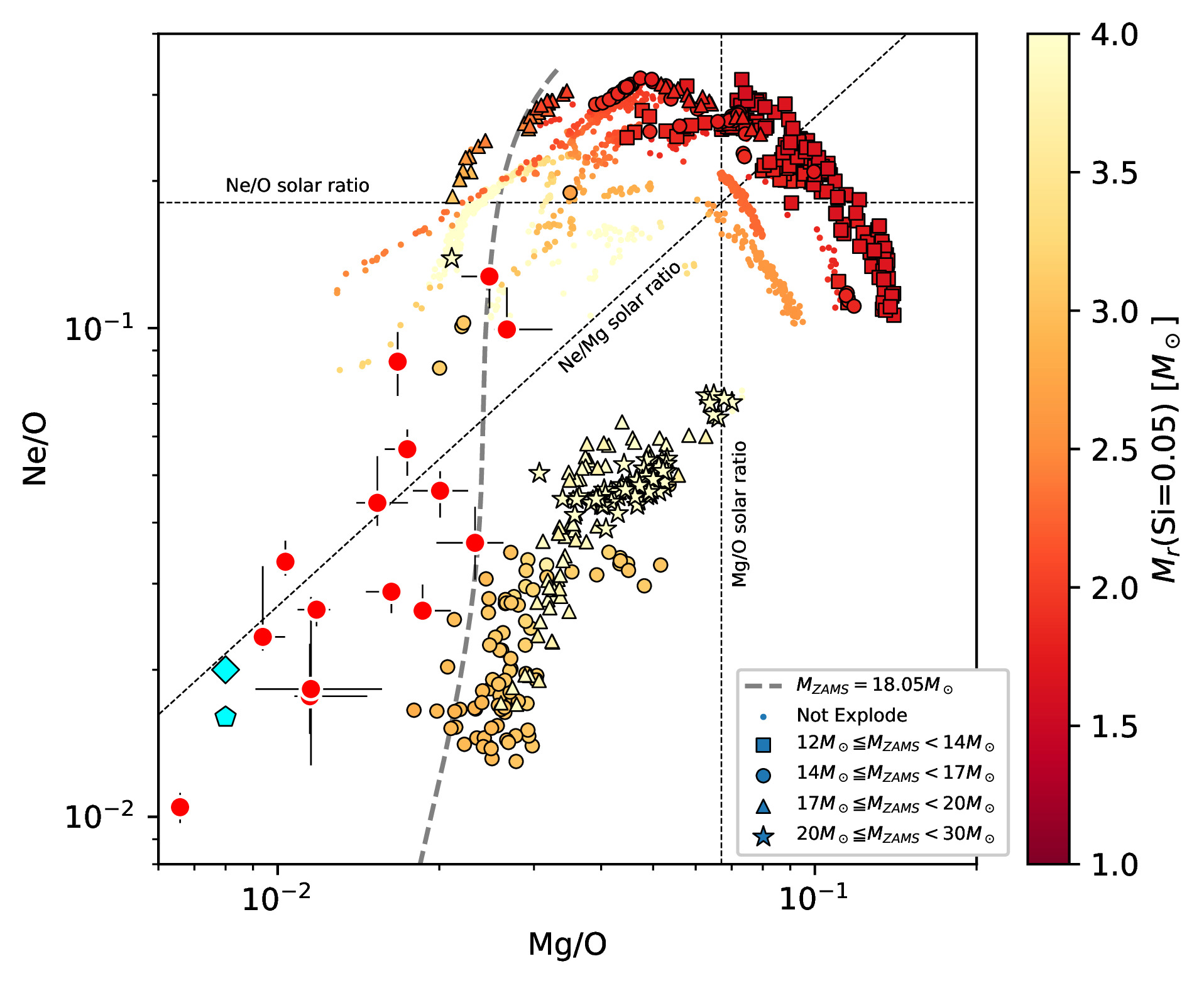}

\caption{The elemental mass ratios (red data points with error bars) in the O-rich regions in Cassiopeia A. Left: A scatter plot between the Si/O and Ne/O mass ratios. The colored data points are derived from the pre-supernova models in \cite{2018ApJ...860...93S}. Right: Same as left figure, but for the Mg/O and Ne/O mass ratios. The cyan diamond and pentagon show the results with the previous X-ray and optical observations \citep{1996A&A...307L..41V,1979ApJ...233..154C}. The gray broken line shows the change in the mass ratios inside the model with $M_{\rm ZAMS} = $ 18.05 $M_\odot$, which is halfway through a shell merger. The radius of the Si-rich region in the progenitor, $M_r$(Si=0.05) (see Figure \ref{fig:shell_pic}), is shown in color. Models that have experienced a shell merger indicate $\gtrsim 3.0 ~M_\odot$ (see also Figure \ref{fig:f2-new} and Figure \ref{fig:S2}). The low Ne/O ratio in Cas~A can only be explained by stars undergoing shell mergers.}\label{fig:f2}
\end{figure*}

\section{Observational support for the shell merger hypothesis} \label{append1}
Previous X-ray observations with different observatories have indicated a low Ne/O ratio for Cassiopeia A \citep{1996A&A...307L..41V,2012ApJ...746..130H}, and this is based on spatially-resolved spectroscopy of the entire remnant. The same conclusion is reached by optical observation \citep{1979ApJ...233..154C}. We compare the observed mass ratios with those in theoretical models in Figure \ref{fig:f2} (see cyan data points), suggesting that such a low Ne/O ratio cannot be explained without shell mergers. Ne, which is the second most abundant element in the largest oxygen layer produced during the evolution of massive stars, is difficult to reduce by one order of magnitude through explosive nucleosynthesis alone \citep{1995ApJS..101..181W}. Thus, we conclude that the shell-merger phenomenon is the most reasonable candidate to explain this feature. Although the observed and theoretical values in Figure \ref{fig:f2} are not in perfect agreement, this could be explained by uncertainties in the use of the O abundance (see also Appendix \ref{append2}) and multidimensional effects in stellar evolution. Regarding the latter, each colored data point in Figure \ref{fig:f2} represents the averaged oxygen layer composition in the 1D pre-supernova models. Our observational measurements, on the other hand, show differences in local regions within a real star. We also note that the convective mixing process itself differs significantly between 1D and 3D \citep[e.g.,][]{2019ApJ...881...16Y}. Therefore, it is possible that these comparison targets have different properties, making complete agreement difficult. For a more detailed understanding, it will be necessary in the future to compare the results with multidimensional calculations of shell mergers that maintain inhomogeneity.

\begin{figure*}[t!]
\centering
\includegraphics[bb=0 0 2601 1269,width=1.0\textwidth]{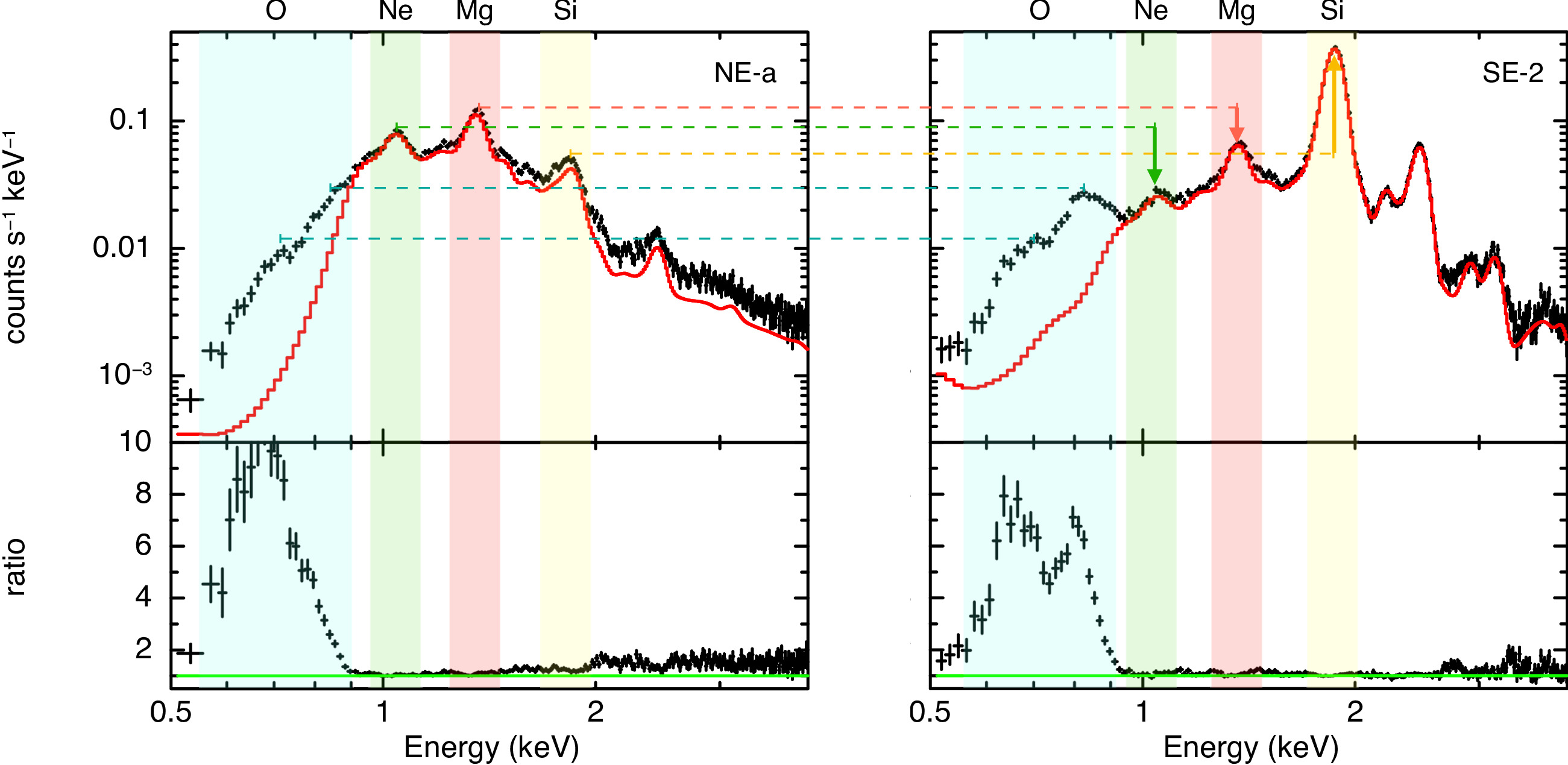}
\caption{Comparison of NE-a and SE-2 spectra. The black data points are the observed X-ray spectra, and the red solid line is the best-fit model excluding all O emissions. The energy bands where the emission lines of O, Ne, Mg, and Si are prominent are highlighted in different colors. The lower panels show the ratios between the model and the data points. In these two spectra, the flux of the O emissions is almost the same, making it easy to compare the differences in the line emissions of Ne, Mg, and Si. The shape of the O-band residuals differs between the two spectra, which is thought to be due to the different contributions of Ne emissions and L-shell emissions from Ar and Ca.
}\label{fig:ONe-OSi}
\end{figure*}

\section{Regarding the use of Ne/Mg and Si/Mg ratios}\label{append2}
Here, we discuss why we used the Ne/Mg and Si/Mg ratios as shown in Fig.~\ref{fig:f2-new} for the main discussion in our study. The main reason for this is the difficulty of evaluating O abundance in observations and theory. As shown in Fig.~\ref{fig:f2}, our measured mass ratios (Ne/O, Si/O, and Mg/O) are systematically lower than those predicted by one-dimensional theoretical models.  Uncertainties in the O abundance measurements and/or in the model calculations complicate direct comparisons between observations and theory.  In contrast, the Ne, Mg, and Si are less affected by these issues, providing a more robust diagnostic of nucleosynthetic yields.

Fig.~\ref{fig:ONe-OSi} shows a spectral comparison between regions NE-a and SE-2. The main challenge in measuring O is the strong interstellar absorption towards this remnant, which makes continuum determination difficult. When all O lines are removed from the model, the residuals shown in the figure become apparent. Since the O emission line intensities in NE-a and SE-2 are nearly identical, spectral differences are easily seen. In NE-a, the high abundance of Ne causes the low ionization Ne He$\alpha$ lines to blend into the O band, affecting the fit. In SE-2, although Ne is less abundant, high Ar and Ca abundances produce L-shell emissions in this energy band. Also, Fe can produce L shell emission that can affect both O and Ne abundance determinations, so it is important to select areas where Fe contribution is low (see section \ref{sec:obs}). Underestimating these contributions would lead to a systematically lower derived O abundance, possibly explaining the deviations seen in Fig.~\ref{fig:f2}. If such an effect also affected the Ne, Mg, and Si measurements, it could alter our conclusions. To test this, we introduced an additional continuum component in the O band and re-fitted the spectrum.

We assess this effect using the NE-a spectrum. First, we added a low-temperature {\tt bremss} component ($k_eT=$0.2 keV) in Xspec; as expected, the O abundance decreased and the Ne/O ratio increased by $\approx$20\%. We then replaced it with a low-temperature {\tt apec} model ($k_eT=$0.2 keV), obtaining essentially the same result. Thus, increasing the continuum contribution in the O band raises the mass ratios relative to O, which may help to account for the trend in Fig.~\ref{fig:f2}. However, the 20\% change is probably too small to resolve all the discrepancies between the models and observations. The multi-dimensional features described below may make comparison with 1D models difficult. Future high-precision X-ray spectroscopy and comparisons with multi-dimensional simulations should help to resolve these issues. On the other hand, the Ne/Mg and Si/Mg ratios remained unchanged ($\lesssim$ 6\%) within 1$\sigma$ statistical errors. Thus, from an observational standpoint, these ratios are more robust than those relative to O. 

Multi-dimensional mixing during the supernova and remnant phases further complicates O‐based comparisons (see also Appendix \ref{append1}).  For example, the stellar O–rich layer contains an outer O/C‐rich region with much lower Ne and Si abundances.  Local mixing with this layer would depress Ne/O and Si/O, producing a trend similar to that in Fig.~\ref{fig:f2}.  Inhomogeneous mixing, which is demonstrated in both explosion and remnant phases \citep[e.g.,][]{1991ApJ...367..619F,2003A&A...408..621K,2010ApJ...714.1371H,2016ApJ...822...22O,2021A&A...645A..66O}, thus complicates O‐based model comparisons.  However, Ne, Mg, and Si coexist within the shell‐merger convection zone, avoiding this bias. Consequently, we conclude that Ne/Mg and Si/Mg ratios represent the least uncertain approach, both observationally and theoretically. 

\begin{figure*}[h!]
\centering
\includegraphics[bb=0 0 1930 1010,width=0.7\textwidth]{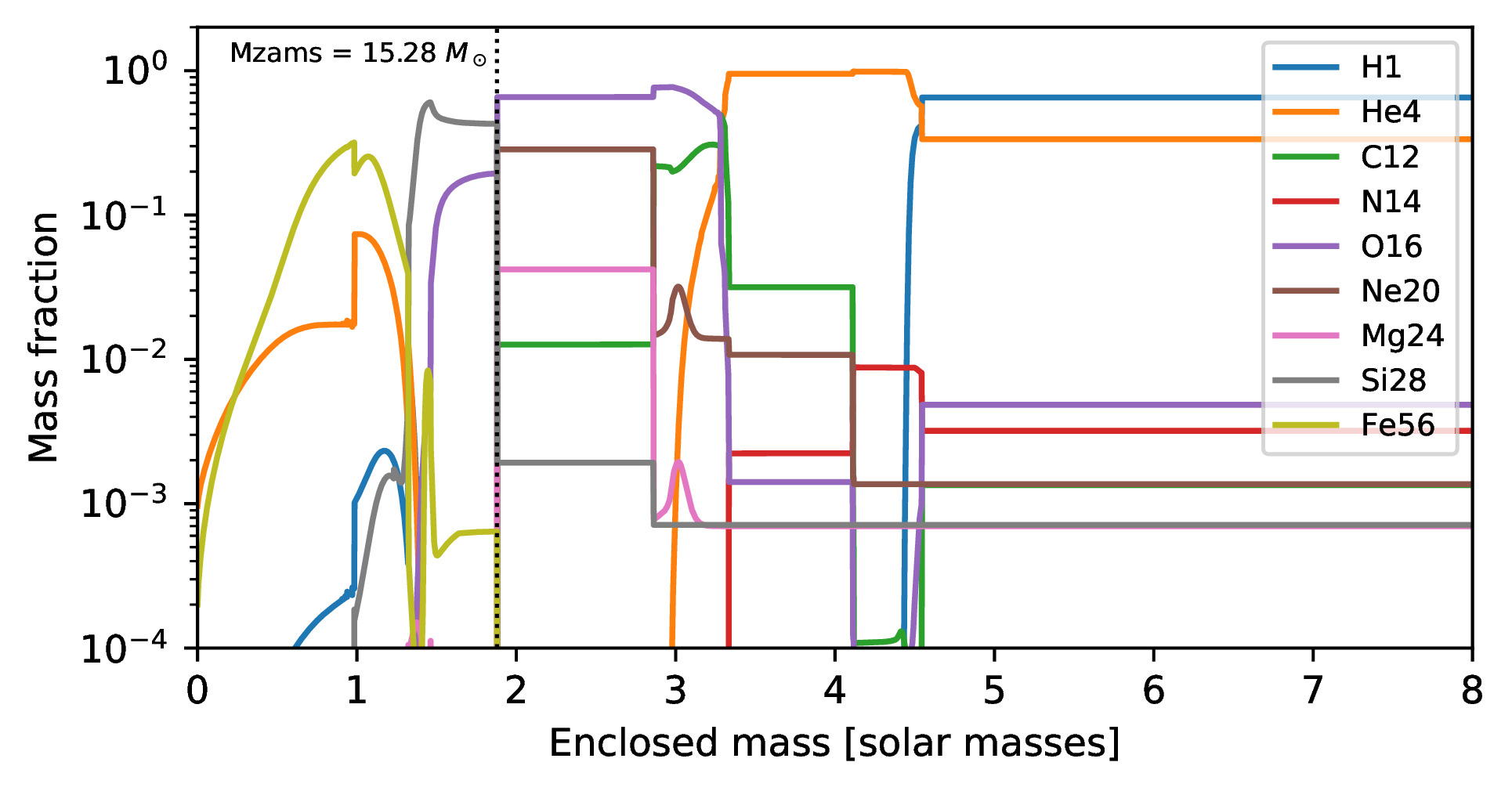}
\includegraphics[bb=0 0 1930 1010,width=0.7\textwidth]{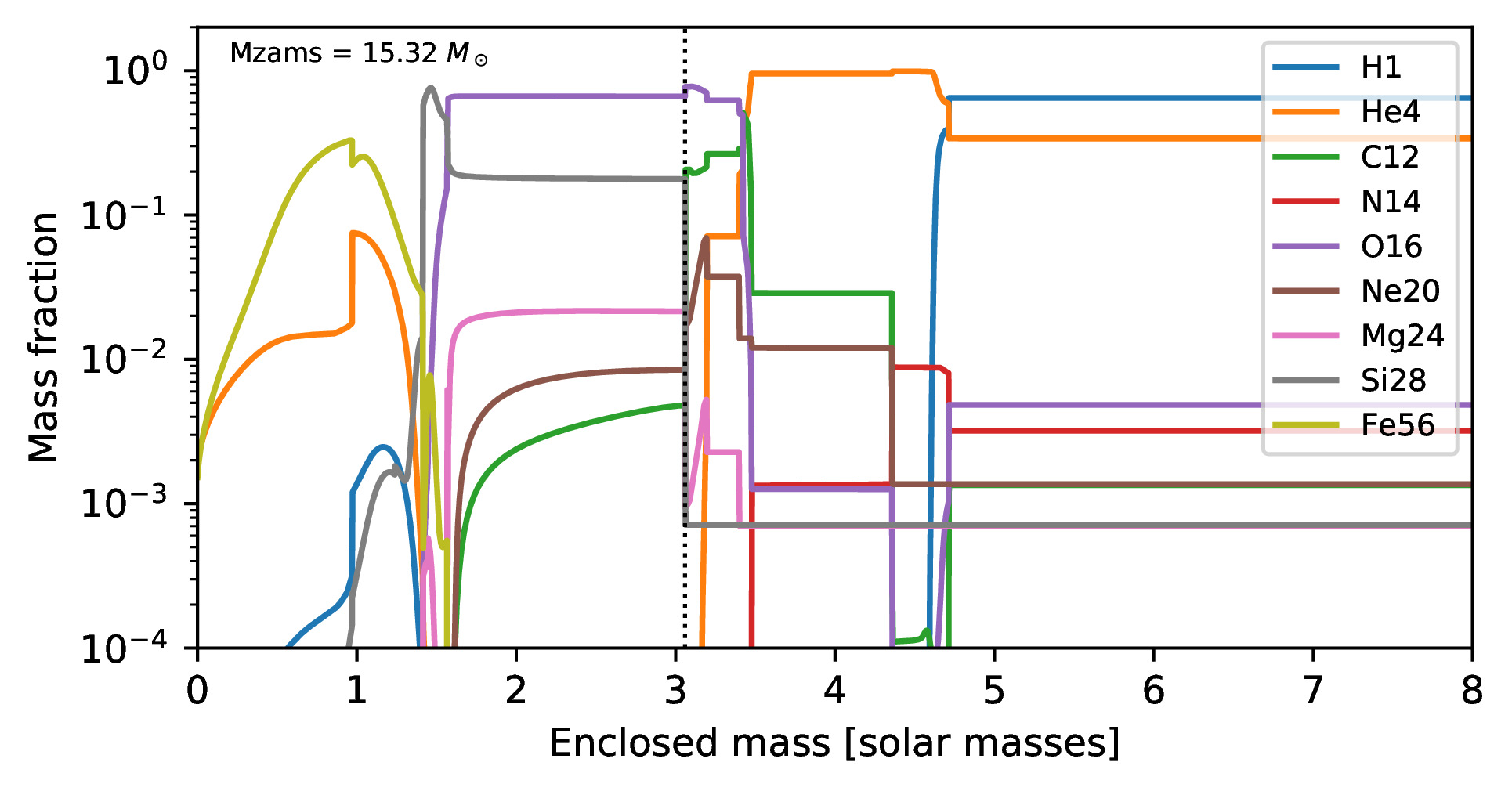}
\includegraphics[bb=0 0 1930 1010,width=0.7\textwidth]{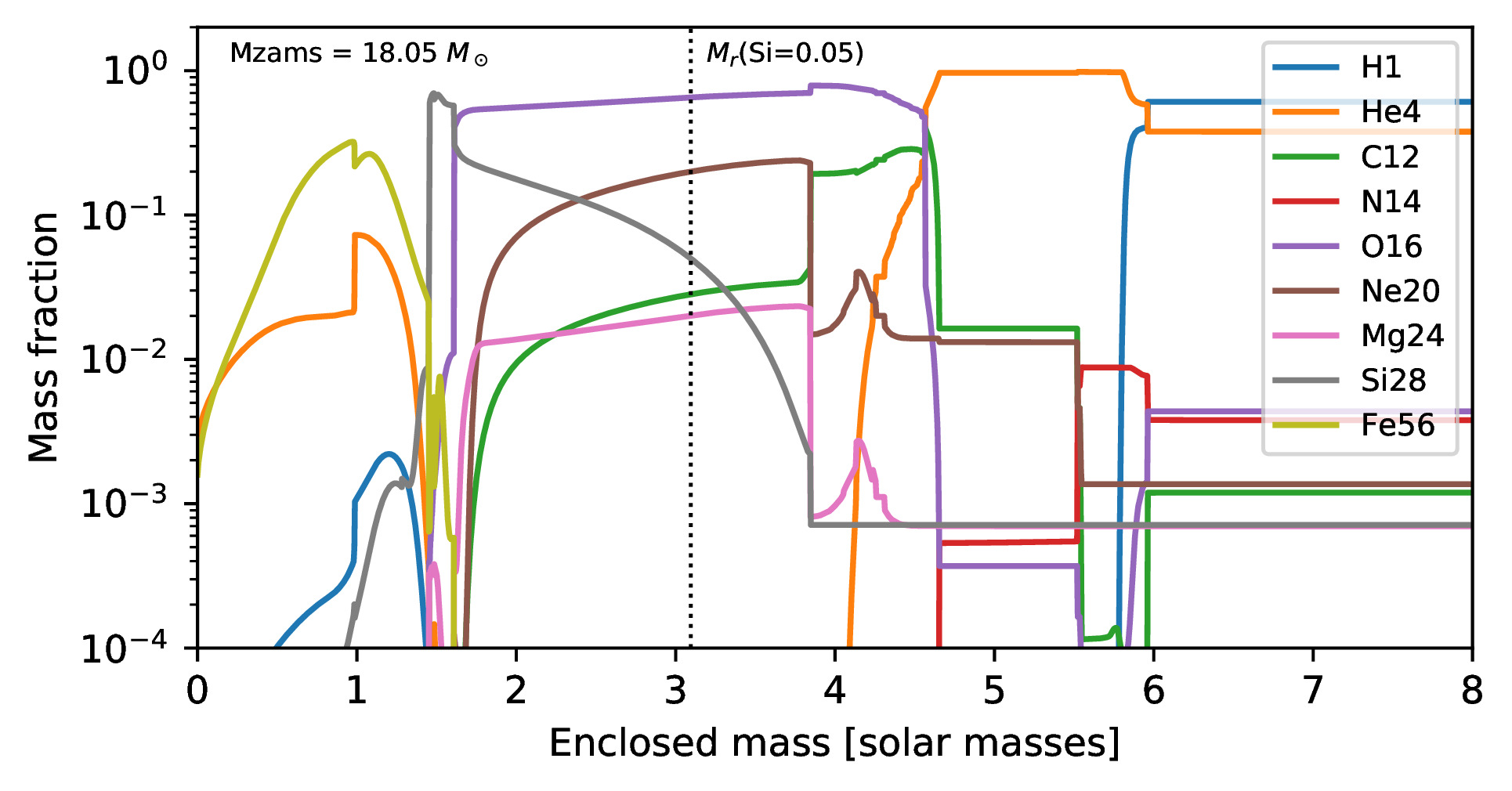}
\caption{Mass fraction profiles for characteristic progenitor models in \cite{2018ApJ...860...93S}. Top: a progenitor model with a standard O-/Ne-rich layer. Middle: a progenitor model that has undergone a shell merger, forming a Ne-poor O-/Si-rich layer. Bottom: a progenitor model with an extended Si-rich layer, possibly undergoing a shell merger. Vertical dotted lines indicate the Si mass radius $M_r$(Si=0.05), where the Si fraction is 0.05. Stars that have experienced a shell merger indicate a larger radius of $\gtrsim 3.0 ~M_\odot$.}\label{fig:S2}
\end{figure*}

\section{Analysis of 1D progenitor models}\label{append3}
To compare with the observational results, we analyzed the 1D pre-supernova models in \cite{2018ApJ...860...93S}. Fig.~\ref{fig:S2} shows a comparison of the mass fraction profiles in some pre-supernova models. In our study, we derived the Si mass radius $M_r$(Si=0.05) to visualize the extended O-/Si-rich layer that is seen when shell mergers occur for the discussion in the main text. The model mass ratios in Fig.~\ref{fig:f2} were calculated using the O-rich layer of pre-supernova models with a high oxygen mass fraction $>$ 0.4.

In Fig. ~\ref{fig:S2}, we see the differences between a typical star and a star that has undergone a shell merger. The 15.28 $M_\odot$ model has a small $M_r$(Si=0.05) of 1.88 $M_\odot$, where the Ne abundance is high in the O-rich layer. This model is thought to be a standard pre-supernova model, where almost 90\% of models in \cite{2018ApJ...860...93S} shows such a Ne/O rich composition. The 15.32 $M_\odot$ model has a large $M_r$(Si=0.05) of 3.05 $M_\odot$, where the Ne-poor and Si-rich composition can be seen in the O-rich layer. We classify these models that have a large Si radius and a low Ne/O ratio into the models that have undergone a shell merger. As shown in Fig.~\ref{fig:f2}, models with a low Ne/O mass ratio of $<$0.1 and a large $M_r$(Si=0.05) of $\gtrsim$ 3.0 $M_\odot$ form a cluster in the scatter plots. These models are thought to have experienced a significant Ne depletion by the shell merger. The 18.05 $M_\odot$ model also has a large $M_r$(Si=0.05) of 3.09 $M_\odot$, where the elemental fractions are gradually changing in the O-rich layer. The gradual change in the elemental fraction suggests extensive, but not complete, convective mixing of the O-rich layer. The change in mass ratios within this O-rich layer is indicated by the gray dashed lines in Fig.~\ref{fig:f2}, where the Mg/O and Ne/O trends follow well those in Cassiopeia A, but the Si/O and Ne/O trends do not.

\begin{table}[h]
\caption{Same as Table~\ref{tab2}, but for two thermal models: {\tt phabs$\times$(nei+nei)}. The abundances tied between the two {\tt vvnei} components.}\label{tab2}
\begin{tabular*}{\textwidth}{@{\extracolsep\fill}lcccc}
\toprule%
Region & NE-a & NE-c & NE-3 & SE-2 \\
\hline
$N_{\rm H}$ [10$^{22}$ cm$^{-2}$]               & 1.34$\pm$0.01 & 1.21$\pm$0.01          & 1.177$^{+0.007}_{-0.003}$       & 1.212$^{+0.006}_{-0.005}$ \\
$kT_{\rm e,1}$ [keV]                            & 4.6$^{+0.3}_{-0.2}$ & 5.9$^{+0.8}_{-0.5}$ & 3.9$^{+0.3}_{-0.2}$          & 3.76$^{+0.02}_{-0.03}$ \\
$kT_{\rm e,2}$ [keV]                            & 0.95$^{+0.03}_{-0.02}$ & 0.88$^{+0.02}_{-0.01}$ & 0.78$\pm$0.01          & 1.30$^{+0.01}_{-0.02}$ \\
(O/H)/(O/H)$_\odot$                                               & 5.6$\pm$0.1    & 27.7$^{+6.5}_{-0.7}$    & 10.8$\pm$0.2                  & 10 (fix)    \\
(Ne/H)/(Ne/H)$_\odot$                                              & 3.6$^{+0.1}_{-0.3}$    & 5.4$^{+0.2}_{-0.4}$ & 1.51$^{+0.03}_{-0.07}$    & 1.13$^{+0.08}_{-0.07}$    \\
(Mg/H)/(Mg/H)$_\odot$                                              & 1.82$^{+0.02}_{-0.11}$ & 4.04$^{+0.76}_{-0.07}$ & 1.95$^{+0.05}_{-0.03}$ & 2.12$^{+0.05}_{-0.07}$ \\
(Si/H)/(Si/H)$_\odot$                                              & 0.54$^{+0.02}_{-0.01}$ & 3.2$\pm$0.5          & 6.2$^{+0.7}_{-0.1}$      & 17.5$^{+0.4}_{-0.2}$  \\
(S/H)/(S/H)$_\odot$                                               & 0.55$\pm$0.04          & 3.2$\pm$0.5       & 4.6$^{+0.1}_{-0.2}$         & 14.1$^{+0.3}_{-0.2}$ \\
(Ca/H)/(Ca/H)$_\odot$ (=Ca)                                        & 0.5$\pm$0.2          & 3.4$\pm$0.8            & 6.8$^{+0.7}_{-0.9}$      & 13.7$^{+0.7}_{-0.6}$   \\
(Fe/H)/(Fe/H)$_\odot$ (=Ni)                                        & 0.13$\pm$0.03        & 0.3$\pm$0.1           & 0.13$^{+0.03}_{-0.02}$    & 0.22$^{+0.04}_{-0.05}$ \\
$n_{\rm e} t_{\rm 1}$  [10$^{9}$ cm$^{-3}$ s] & 9.7$\pm$0.3            & 7.8$\pm$0.3           & 5.6$\pm$0.2               & 9.8$^{+0.8}_{-0.5}$            \\
$n_{\rm e}t_{\rm 2}$  [10$^{10}$ cm$^{-3}$ s]& 19.3$\pm$0.05    & 20$\pm$2                     & 9.2$^{+0.1}_{-0.3}$       & 6.3$^{+0.2}_{-0.3}$    \\
Redshift $z_1$ [$\times 10^{-2}$]       & $-$1.994$\pm$0.004    & $-$1.089$^{+0.001}_{-0.002}$  & $-$1.43$^{+0.14}_{-0.03}$ & $-$2.39$^{+0.02}_{-0.01}$   \\
Redshift $z_2$ [$\times 10^{-2}$]        & $-$0.96$^{+0.03}_{-0.04}$    & $-$0.794$\pm$0.001    & $-$0.031$^{+0.052}_{-0.005}$ & $-$1.824$^{+0.051}_{-0.004}$   \\
$\chi^2$/d.o.f                           & 312.66/297             & 296.72/260                  & 291.96/239                & 320.67/265             \\
\hline
\end{tabular*}
\end{table}

\begin{table}[h]
\caption{Elemental mass ratios in each O-rich region. The errors are 1$\sigma$ level.}\label{tab3}
\begin{tabular*}{\textwidth}{@{\extracolsep\fill}lccccc}
\toprule%
    & $M_{\rm Ne}/M_{\rm O}$ & $M_{\rm Mg}/M_{\rm O}$ & $M_{\rm Si}/M_{\rm O}$ &$M_{\rm Si}/M_{\rm Mg}$ & $M_{\rm Ne}/M_{\rm Mg}$\\
\hline
NE-a   & (1.3$^{+0.1}_{-0.2}$)$\times10^{-1}$& (2.5$^{+0.1}_{-0.3}$)$\times10^{-2}$& (7.7$^{+0.7}_{-1.0}$)$\times10^{-3}$ & (3.1$^{+0.3}_{-0.5}$)$\times10^{-1}$ & 5.1$^{+0.4}_{-0.9}$ \\%OMgrich_Sipoor_1
NE-b   & (4.4$^{+1.1}_{-0.5}$)$\times10^{-2}$& (1.5$^{+0.2}_{-0.1}$)$\times10^{-2}$& (1.1$^{+0.2}_{-0.1}$)$\times10^{-2}$  & (6.9$^{+1.5}_{-0.7}$)$\times10^{-1}$ & 2.9$^{+0.8}_{-0.4}$\\%OMgrich_Sipoor_3
NE-c   & (5.7$^{+0.6}_{-0.7}$)$\times10^{-2}$& (1.7$^{+0.1}_{-0.2}$)$\times10^{-2}$& (1.4$\pm$0.1)$\times10^{-2}$  & (8.1$^{+0.6}_{-1.1}$)$\times10^{-1}$ & 3.2$^{+0.4}_{-0.5}$\\%OMgrich_Sipoor_2
NE-d   & (8.5$^{+1.3}_{-1.3}$)$\times10^{-2}$& (1.6$\pm$0.1)$\times10^{-2}$& (1.8$^{+0.2}_{-0.1}$)$\times10^{-2}$  & 1.1$\pm$0.1 & 5.1$\pm$0.8\\%OMgrich_Sipoor_6
NE-e   & (9.9$^{+2.2}_{-0.5}$)$\times10^{-2}$& (2.7$^{+0.6}_{-0.1}$)$\times10^{-2}$& (2.0$^{+0.4}_{-0.1}$)$\times10^{-2}$ & (7.7$^{+2.3}_{-0.6}$)$\times10^{-1}$ & 3.7$^{+1.1}_{-0.3}$ \\%OMgrich_Sipoor_4
NE-f   & (3.3$^{+0.3}_{-0.2}$)$\times10^{-2}$& (1.03$\pm$0.05)$\times10^{-2}$& (1.17$\pm$0.05)$\times10^{-2}$  & 1.1$\pm$0.1 & 3.2$^{+0.2}_{-0.3}$\\%OMgrich_Sipoor_7
SE-a   & (4.6$^{+0.4}_{-0.5}$)$\times10^{-2}$& (2.0$^{+0.3}_{-0.2}$)$\times10^{-2}$& (2.8$^{+0.2}_{-0.3}$)$\times10^{-2}$  & 1.4$^{+0.1}_{-0.2}$ & 2.3$\pm$0.4\\%OMgrich_Sipoor_5
NE-1   & (1.05$\pm$0.01)$\times10^{-2}$& (6.6$\pm$0.2)$\times10^{-3}$& (4.3$^{+0.2}_{-0.1}$)$\times10^{-2}$  & 6.6$^{+0.3}_{-0.2}$ & 1.6$\pm$0.1\\%OSi-rich_small_4
NE-2   & (2.6$^{+0.1}_{-0.2}$)$\times10^{-2}$& (1.1$\pm$0.1)$\times10^{-3}$& (7.4$^{+0.2}_{-0.6}$)$\times10^{-2}$   & 6.3$^{+0.4}_{-0.7}$ & 2.2$^{+0.1}_{-0.2}$\\%OSi-rich_small_10
NE-3   & (3.6$^{+0.7}_{-0.6}$) $\times10^{-2}$& (2.3$^{+0.3}_{-0.4}$)$\times10^{-2}$& (6.3$^{+1.0}_{-0.7}$)$\times10^{-2}$  & 2.7$^{+0.6}_{-0.5}$ & 1.6$\pm$0.3\\%Orich_small_1
NE-4   & (2.9$^{+0.1}_{-0.3}$)$\times10^{-2}$& (1.63$^{+0.04}_{-0.17}$)$\times10^{-2}$& (1.02$^{+0.02}_{-0.09}$)$\times10^{-1}$  & 6.3$^{+0.9}_{-0.2}$ & 1.8$^{+0.1}_{-0.2}$\\%OSi-rich_small_3
SE-1   & (2.3$^{+0.9}_{-0.1}$)$\times10^{-2}$& (9.3$^{+0.9}_{-0.4}$)$\times10^{-3}$& (3.8$^{+0.6}_{-0.1}$)$\times10^{-2}$  & 4.1$^{+0.7}_{-0.2}$ & 2.5$^{+1.0}_{-0.2}$\\%Orich_small_3
SE-2   & (2.6$^{+0.3}_{-0.1}$)$\times10^{-2}$& (1.9$^{+0.2}_{-0.1}$)$\times10^{-2}$& (1.64$^{+0.21}_{-0.04}$)$\times10^{-1}$  & 8.8$^{+1.6}_{-0.3}$ & 1.4$^{+0.3}_{-0.1}$\\%OSi-rich_small_1
SE-3   & (1.8$^{+1.0}_{-0.6}$) $\times10^{-2}$& (1.1$^{+0.4}_{-0.2}$)$\times10^{-2}$& (2.2$^{+0.8}_{-0.4}$)$\times10^{-2}$  & 1.9$^{+0.9}_{-0.5}$ & 1.6$^{+1.0}_{-0.6}$\\%Orich_small_2
SE-4   & (1.8$^{+0.5}_{-0.3}$)$\times10^{-2}$& (1.1$^{+0.3}_{-0.7}$)$\times10^{-2}$& (6.0$^{+1.7}_{-2.0}$)$\times10^{-2}$  & 5.2$^{+2.1}_{-1.8}$ & 1.5$^{+0.6}_{-0.3}$\\%OSi-rich_small_2
\hline
\end{tabular*}
\end{table}

\bibliography{sample7}{}
\bibliographystyle{aasjournalv7}



\end{document}